\documentclass[preprintnumbers,amsmath,amssymb,floatfix,10pt,prd,onecolumn,
superscriptaddress,nofootinbib]{revtex4}
\usepackage{latexsym}
\usepackage{epsfig}
\usepackage{epstopdf}
\usepackage{graphicx}
\usepackage{amssymb}
\usepackage{amsmath}
\usepackage{dcolumn}
\usepackage{bm}
\usepackage{color,float}
\usepackage{comment}
\begin{document}
\title{Cosmic evolution in the background of non-minimal coupling in $f(R,T,R_{\mu\nu}T^{\mu\nu})$ Gravity}
\author{M. Zubair}
\email{mzubairkk@gmail.com;drmzubair@ciitlahore.edu.pk}\affiliation{Department
of Mathematics, COMSATS Institute of Information Technology Lahore,
Pakistan}
\author{M. Zeeshan}
\email{m.zeeshan5885@gmail.com}\affiliation{Department
of Mathematics, COMSATS Institute of Information Technology Lahore,
Pakistan}

\begin{abstract}
An accelerated expansion phase is being experienced by the universe due to the presence of an unknown energy component known as dark energy (DE). To find out the cosmic evolution scientists ever tried to modify Einstein's gravitational theory and its unexplored parts. We also look forward to address the same problem with a different approach based on interaction between matter and geometry. For this purpose we consider $f(R,T,Q)$ modified theory (where $R$ is the Ricci Scalar, $T$ is the trace of energy-momentum tensor (EMT) $T_{uv}$ and $Q=R_{uv}T^{uv}$ is interaction of EMT $T_{\mu\nu}$ and Ricci Tensor $R_{uv}$). We formulate modified field equations in the background of Friedmann-Lema$\hat{i}$tre-Robertson-Walker (FLRW) model which is defined as $ds^2=dt^2-a(t)^2(dx^2+dy^2+dz^2 )$, where $a(t)$ represents the scale factor. In this formalism energy density is found using covariant divergence of modified field equations. $\rho$ involves a contribution from non-minimal matter geometry coupling which helps to study different cosmic eras based on equation of state (EOS). Furthermore, we apply the energy bounds to constrain the model parameters establishing a pathway to discuss the cosmic evolution for best suitable parameters in accordance with recent observations.
\end{abstract}

\date{\today}

\maketitle

\section{introduction}

Currently our universe is experiencing an accelerated expansion phase and multiple astrophysical researches have been conducted to observe this cosmic scenario. It is highly assumed and considered that this cosmic acceleration is the consequence of an anonymous energy named as DE \cite{1}. Antagonistic to the gravitational pull, the DE is expanding the universe by having a negative pressure which is completely opposite to the ordinary matter. Many attempts have been made to unveil the reason for accelerated cosmic expansion. The major finding \cite{2} enlists DE as the major candidate with overall contribution of 68.3\%,\ the other significant 26.8\%\ contribution is from Dark matter despite its elusive and un-explored nature. Baryon, is the major part of visible cosmos which accounts for 4.9\%\ among cosmic ingredients. Despite tremendous researches and observations, late time cosmic acceleration is still a significant as well as challenging area for cosmologists. However, attention is attached to the confirmation through measurements from temperature anisotropies of the existence of DE as puzzling cosmic ingredient with reference to cosmic acceleration by cosmic microwave background radiations (CMBR) \cite{2}, baryon acoustic oscillations (BAO) \cite{3}, large scale structure (LSS) \cite{4}, weak lensing \cite{5} and most recent plank's data \cite{6}. Furthermore, to describe the nature of DE several theoretical models are proposed like phantom \cite{7}, quintessence \cite{8} and fluids with anisotropic equation of state (EoS) \cite{9}. In $\Lambda$ cold dark matter ($\Lambda$CDM) model, the role of DE in GR is played by $\Lambda$. Yet the origin of cosmological constant $\Lambda$ is still under question and $\Lambda$ has two well-known problems known as coincidence and fine-tuning. To express the characteristics of DE, the EoS is proposed which is defined as $\omega_{DE}=\frac{p_{DE}}{\rho_{DE}}$ (the ratio of the pressure to the energy density of DE). The equation of state is evaluated by considering that universe is isotropic and homogeneous, and taking the  FLRW space-time at the background. $\omega_{DE}$ is a constant and equal to -1 in $\Lambda$CDM model whereas in quintessence model $\omega_{DE}$ is dynamical quantity and $-1<\omega_{DE}<\frac{-1}{3}$. Moreover $\omega_{DE}$ varies with time and $\omega_{DE}<-1$ in phantom model. It concludes that different model descriptions such as fluid description and the description of a scalar field theory might describe cosmic picture. DE models can also be constructed by modification of Einstein Hilbert action that further leads to the modified gravity models. The preliminary step is to substitute Einstein Hilbert term by scalar curvature and it results in the formation of $f(R)$ gravity \cite{10}. In this theory, the general non-linear function $f$ depends on the Ricci scalar and if we replace this generic function $f$ by $f\equiv R-2\Lambda$ then we will get the classic $\Lambda$ cold dark matter ($\Lambda$CDM) model. This theory is also interesting due to the fact that for a specific BD parameter \cite{11} it develops correspondence with the BD theory that include nonminimal coupling between scalar field and geometry. This coupling is also further constructed in $f(R)$ theory \cite{12,13}.
Bertolami et al. \cite{13} gave another direction to $f(R)$ gravity where they coupled matter Lagrangian density $\mathcal{L}_m$ with the Lagrangian as a function of scalar curvature. In \cite{14}, authors developed the equivalence of scalar tensor theories with this theory which involves non-minimal scalar curvature term and a non-minimum coupling of the scalar curvature and matter.
This nonminimal coupling further lead to non-conserved matter energy-momentum tensor(EMT) which results in non-geodesic motion of test particles \cite{15}. In \cite{16}, Harko generalized this non-minimal coupling by introducing a function of matter Lagrangian. Later Wu \cite{17} further extended this work by studying few forms of curvature components and forming the thermodynamic laws. Harko along with the contributions of Lobo \cite{18} proposed another induced form of $f(R)$ by involving curvature matter coupling incorporating matter langrangian $\mathcal{L}_m$ and defined generic function $f(R,\mathcal{L}_m)$. In \cite{zfRLM}, Sharif and Zubair discussed the non-equilibrium thermodynamics in $f(R,\mathcal{L}_m)$ gravity, and develop constraints on two specific
gravitational models $f(R,\mathcal{L}_m)=\lambda{\exp}\left(\frac{1}{2\lambda}R+\frac{1}{\lambda}\mathcal{L}_m\right)$
and $f(R,\mathcal{L}_m)=\alpha{R}+\beta{R}^2+\gamma{\mathcal{L}_m}$ to secure the validity of GSLT in this theory. In \cite{baha}, authors presented the torsion-matter coupling and inclusion of boundary term to discuss different cosmic issues. 

The selection of matter Lagrangian density has an issue in modified theories, specifically for those which involve nonminimal coupling with matter Lagrangian. For the natural conservation of matter we are restricted to take matter Lagrangian as $\mathcal{L}_m=p$ then extra force will be vanished \cite{19} or for the sake of effective nonminimal coupling we can also take $\mathcal{L}_m=-\rho$ \cite{20}. Alternative method to modify the Lagrangian of Einstein's equations is to take the function which depends on trace $T$ of the EMT \cite{21}, such that $\Lambda$CDM model can be taken of the form $R+2\Lambda(T)$. Finally, by using this idea Harko et al. \cite{22} proposed the extension of $f(R)$ gravity by replacing the function $f$ with the new dependent parameters $R$ (Ricci scalar) and $T$ (trace of energy-momentum tensor) and nonminimal coupling between matter and geometry allowed in this astonishing theory. The coinciding with constructive geometry matter coupling shows the deviation of test particles from geodesic motion which ruled to additional force as proposed in different theories \cite{13,16,19,22}.\\
Due to remarkable growing interest in this theory the efficacy of laws of thermodynamics in $f(R,T)$ gravity have been studied by Sharif and Zubair \cite{23} and it is concluded that the equilibrium picture of thermodynamics cannot be achieved due to matter geometry interaction. Attempts to reconstruct $f(R, T)$ Lagrangian has also been made under various considerations like the family of holographic DE models by supposing the FLRW universe \cite{24}, considering an auxiliary scalar field \cite{25} and anisotropic solutions \cite{26}. Jamil et.al \cite{27} worked on the reconstruction of cosmological models and they showed that the dust fluid reproduce $\Lambda$CDM, Einstein static universe and de sitter Universe. Alvarenga et al. \cite{28} discussed the development of matter density perturbations in this theory and they presented the required constraints to get the standard continuity equation in $f(R,T)$ gravity. On the other hand, Sharif and Zubair \cite{29} reconstructed cosmological models by applying additional constraints for the conserved EMT and studied the stability of the constructed models. Furthermore, the dynamical systems in $f(R,T)$ gravity were explored by Shabani and Farhoudi \cite{30} that resulted in the development of a vast scale of passable cosmological solutions. Other cosmic issues including compact stars, wormholes and gravitational instability of collapsing stars have been discussed in literature \cite{fRTlit}.

Lately, the non-minimal coupling of the EMT and Ricci tensor is introduced, resulting in the modified yet more complicated theory known as $f(R,T,Q)$ gravity \cite{31,32}. Due to complicated nonminimal matter-geometry coupling EMT is generally non-conserved and additional force is there. Therefore, it proposes a vast range to explore different cosmic features as thermodynamics properties have already been studied by Sharif and Zubair \cite{33} and they \cite{34} also discussed the energy conditions for particular models of $f(R,T,Q)$ gravity. They found that the nonminimal coupling becomes the reason of the deviation of test particles from geodesic motion and that gives strength to the non-equilibrium representation of thermodynamics. This induced the idea that the validity of generalized second law of thermodynamics (GSLT) in an expanding universe might lead to the thermal equilibrium in future. E.H.Baffou et al. \cite{35} discussed the stability of de-sitter and power law solution by using perturbation scheme for particular models.
In this paper we are interested to discus the cosmological evolution in $f(R,T,Q)$ theory, which is based on more general matter-geometry coupling. We pick a particular model of the form $f(R,T,Q)=R+\alpha{Q}+\beta{T}$, and solve the matter conservation equation to find the explicit expression of energy density. Evolution of EoS parameter $\omega_{eff}$ and deceleration parameter is discussed employing the power law cosmology. This manuscript is organized as follows: In Sec. II, we briefly introduce $f(R,T,Q)$ theory and present the general formalism of field equations for a FLRW cosmology. Section III is devoted to particular model in this theory where, we present the expressions for $\rho_{eff}$, $p_{eff}$ and $\omega_{eff}$. In section IV we constrain the model parameters using the energy bounds. Section V, concludes our discussion.

\section{$f(R,T,Q)$ Gravity}

The $f(R,T,Q)$ gravity is the most generalized gravity among other modified gravities like $f(R)$ and $f(R,T)$ and this theory is very effective for nonminimal coupling between matter and geometry. The action of this complicated theory takes the following form \cite{31,32}
\begin{equation}\label{1}
\mathcal{A}=\frac{1}{2{\kappa}^2}\int{dx^4\sqrt{-g}\left[f(R,T,R_{\mu\nu}T^{\mu\nu})
+\mathcal{L}_{m}\right]},
\end{equation}
where $\kappa^2=8\pi G$, $f(R,T,Q)$ is a general function which depends on three components, the Ricci scalar $R$, trace of the EMT $T$, product of the EMT $T^{\mu\nu}$ to Ricci tensor $R_{\mu\nu}$,
and $\mathcal{L}_{m}$ shows the matter Lagrangian. The EMT for matter is defined as
\begin{equation}\label{2}
T_{{\mu}{\nu}}=-\frac{2}{\sqrt{-g}}\frac{\delta(\sqrt{-g}
{\mathcal{\mathcal{L}}_{m}})}{\delta{g^{{\mu}{\nu}}}}.
\end{equation}
If the matter action depends only on the metric tensor other than on its
derivatives then the EMT yields
\begin{equation}\label{3}
T_{{\mu}{\nu}}=g_{{\mu}{\nu}}\mathcal{L}_{m}-\frac{2{\partial}
{\mathcal{L}_{m}}}{\partial{g^{{\mu}{\nu}}}}.
\end{equation}
The field equations in $f(R,T,Q)$ gravity can be found by
varying the action (\ref{1}) with respect to $g_{\mu\nu}$ as
\begin{eqnarray}\nonumber
&&R_{{\mu}{\nu}}f_{R}-\{\frac{1}{2}f-\mathcal{L}_{m}f_T-\frac{1}{2}\nabla_\alpha
\nabla_\beta(f_QT^{\alpha\beta})\}g_{{\mu}{\nu}}+(g_{{\mu}{\nu}}
{\Box}-{\nabla}_{\mu}{\nabla}_{\nu})f_{R}+\frac{1}{2}
\Box(f_QT_{{\mu}{\nu}})+2f_QR_{\alpha(\mu}T^\alpha_{\nu)}-\nabla_\alpha\nabla
_{(\mu}[T^\alpha_{\nu)}f_Q]\\\label{4}
&&-G_{\mu\nu}\mathcal{L}_{m}f_Q-2\left(f_Tg^{\alpha\beta}
+f_QR^{\alpha\beta}\right)\frac{\partial^2\mathcal
{L}_{m}}{\partial{g}^{\mu\nu}\partial{g}^{\alpha\beta}}=(1+f_T+\frac{1}
{2}R{f}_Q)T_{\mu\nu}.
\end{eqnarray}
The subscripts shows the derivatives with respect to $R,T,Q$, and box function defined as ${\Box=\nabla}^{\beta}{\nabla}_{\beta}$, ${\nabla}_{\mu}$ represent covariant derivative. If we will choose the particular form of Lagrangian then Equation (\ref{4}) can be shifted towards the well known field equations in $f(R)$ and $f(R,T)$ theories.
The field equation (\ref{4}) can be rewritten into the form of effective Einstein field equation (EFE) as
\begin{equation}\label{5}
G_{\mu\nu}=R_{{\mu}{\nu}}-\frac{1}{2}Rg_{{\mu}{\nu}}=T_{{\mu}{\nu}}^{eff}.
\end{equation}
This effective form of EFE is similar to GR's standard field equations. Here
$T_{{\mu}{\nu}}^{eff}$, the effective EMT in
$f(R,T,Q)$ gravity is found to be as
\begin{eqnarray}
&&\nonumber
{T}_{{\mu}{\nu}}^{eff}=\frac{1}{f_{R}-f_Q\mathcal{L}_{m}}\left[(1+f_T+
\frac{1}{2}R{f}_Q)T_{\mu\nu}+\{\frac{1}{2}(f-R{f}_R)-\mathcal{L}_{m}f_T
\frac{1}{2}\nabla_\alpha
\nabla_\beta(f_QT^{\alpha\beta})\}g_{{\mu}{\nu}}-(g_{{\mu}{\nu}}
{\Box}-{\nabla}_{\mu}{\nabla}_{\nu})f_{R}\right.\\&&\left.-\frac{1}{2}
\Box(f_QT_{{\mu}{\nu}})2f_QR_{\alpha(\mu}T^\alpha_{\nu)}+
\nabla_\alpha\nabla_{(\mu}[T^\alpha_{\nu)}f_Q]+2\left(f_Tg^{\alpha\beta}
+f_QR^{\alpha\beta}\right)\frac{\partial^2\mathcal{L}_{m}}{\partial{g}^{\mu\nu}
\partial{g}^{\alpha\beta}}\right].
\end{eqnarray}
Applying the covariant divergence to the field equation (\ref{4}),
we get
\begin{eqnarray}
&&\nonumber
\nabla^{\mu}T_{\mu\nu}=\frac{2}{2(1+f_T)+R{f}_Q}
\left[\nabla_\mu(f_QR^{\alpha\mu}T_{\alpha\nu})+\nabla_\nu(\mathcal{L}_mf_T)
-\frac{1}{2}(f_Q{R}_{\sigma\zeta}f_T{g}_{\sigma\zeta})
\nabla_\nu{T}^{\sigma\zeta}-G_{\mu\nu}\nabla^\mu(f_Q\mathcal{L}_m)\right.\\\label{cons}&&\left.-\frac{1}{2}
\left[\nabla^\mu(R{f}_Q)+2\nabla^\mu{f}_T\right]T_{\mu\nu}\right].
\end{eqnarray}
It is important to see that any modified theory which involve nonminimal coupling between geometry and matter does not obey the ideal continuity equation. This complicated theory $f(R,T,Q)$ also involves this type of nonminimal coupling so it also deviate from standard behavior of continuity equation. Here, non-minimal coupling between matter and geometry induces extra force acting on massive particles, whose equation of motion is given by \cite{32}
\begin{eqnarray}\nonumber
\frac{d^2x^\lambda}{ds^2}+\Gamma^\lambda_{\mu\nu}u^\mu u^\nu=f^\lambda,
\end{eqnarray}
where
\begin{eqnarray}\nonumber
&&f^\lambda=\frac{h^{\lambda\nu}}{(\rho+p)(1+2f_T+Rf_{RT})}\Big[(f_T+Rf_{RT})\nabla_\nu\rho-(1+3f_T)\nabla_\nu p-(\rho+p)f_{RT}R^{\sigma\rho}(\nabla_\nu h_{\sigma\rho}-2\nabla_\rho h_{\sigma\nu})\\\nonumber&&-f_{RT}R_{\sigma\rho}h^{\sigma\rho}\nabla_\nu(\rho+p)\Big].
\end{eqnarray}
It has been found that the impact of non-minimal coupling is always present independent of the choice matter Lagrangian, the extra force does not vanish even with the Lagrangian $\mathcal{L}_m=p$ as compared to the results presented in \cite{coupling}. In \cite{32}, authors also presented the Lagrange multiplier approach and found the conservation of matter EMT. Moreover, if one eliminates the dependence of $Q$, it results in divergence equation of $f(R,T)$ theory as given below
\begin{equation}\nonumber
\nabla{^{\alpha}}T_{\alpha\beta}=\frac{f_T}{1-f_T}\left[(\Theta_{\alpha\beta}+T_{\alpha\beta})\nabla{^{\alpha}}ln
f_T-\frac{1}{2}g_{\alpha\beta}\nabla{^{\alpha}}T+\nabla^{^{\alpha}}\Theta_{\alpha\beta}\right].
\end{equation}
In \cite{28}, Alvarenga et al.shown that choice of a specific model within
these theories can guarantee the conservation of EMT and continuity equation is valid for the
model $f(R,T)=f_1(R)+f_2(T)$, where $f_2(T)=\alpha T^{\frac{1+3\omega}{2(1+\omega)}}+\beta$. In this manuscript, we are interested to evaluate the role of non-minimal coupling in cosmic evolution so we opt the nonconserved dynamical equation and evaluate necessary parameters.

We consider the isotropic and homogenous flat FLRW metric defined as
\begin{equation}\nonumber
ds^{2}=dt^2-a^2(t)(dx^2+dy^2+dz^2),
\end{equation}
where $a(t)$ represents the scale factor. The effective energy density and pressure for this metric is found to be the components of $T_{\mu\nu}^{eff}$, which assumes the form of perfect fluid as
\begin{equation}
T_{\mu\nu}=(p+\rho)u_\mu u_\nu-pg_{\mu\nu}
\end{equation}
where $p$ represent pressure, $\rho$ for proper density and $u_\mu$ is for 4-velocity. In FLRW background, $\rho_{eff}$ and $p_{eff}$ can be found as
\begin{eqnarray}&&\nonumber
\rho_{eff}=\frac{1}{f_{R}-f_Q\mathcal{L}_{m}}\left[\rho+(\rho
-\mathcal{L}_m)f_T+\frac{1}{2}(f-R{f}_R)-3H\partial_t{f}_R-\frac{3}{2}(3H^2-
\dot{H})\rho{f}_Q-\frac{3}{2}(3H^2+\dot{H})pf_Q+\right.\\&&\left.\frac{3}{2}H\partial_t
[\left(p-\rho)f_Q\right]\right],
\\\nonumber &&
{p}_{eff}=\frac{1}{f_{R}-f_Q\mathcal{L}_{m}}\left[p+(p
+\mathcal{L}_m)f_T+\frac{1}{2}(R{f}_R-f)+\frac{1}{2}(\dot{H}+3H^2)(\rho-p){f}_Q
+\partial_{tt}{f}_R+2H\partial_tf_R+\frac{1}{2}\partial_{tt}[(\rho-p)f_Q]
\right.\\&&\left.+2H\partial_t
[\left(\rho+p){f}_Q\right]\right],
\end{eqnarray}
where $R=-6(\dot{H}+2H^2), ~H=\frac{\dot{a}}{a}$ is for Hubble
parameter and upper dot for the time derivative. Here, we ignored those terms which involved the second derivative of matter Lagrangian with respect to $g_{\mu\nu}$.
In the case of perfect fluid the matter Lagrangian can either be $\mathcal{L}_m=\rho$ or $\mathcal{L}_m=-p$.

\section{$f(R,T,Q)=R+\alpha{Q}+\beta{T}$ Gravity}

We are interested to explore the cosmic evolution using matter conservation equation of more generic modified theory. Here, we will set $\mathcal{L}_m=\rho$ and we will take the simplest model
$f(R,T,Q)=R+\alpha{Q}+\beta{T}$
where $\alpha$, $\beta$ are coupling parameters. In this model, choice of $\alpha=0$ results in minimal coupling of the form $f(R,T)=R+\beta T$ \cite{22}, such model has been widely studied in the formalism of $f(R,T)$ gravity (for review see \cite{fRTlit}). Moreover, the choice of $\alpha=\beta=0$, results in Eisntein's formalism of GR.

For a flat FLRW universe, the non-zero components of FLRW equation for ${p}_{eff}={p}+{p}_{DE}$ and
${\rho}_{eff}={\rho}+{\rho}_{DE}$ are
\begin{eqnarray}\nonumber
3{H}^{2}={\rho}_{eff},\\-2{\dot{H}}-3{H^{2}}={P}_{eff},
\end{eqnarray}
where dots being time derivative and components of $\rho_{DE}$ and $p_{DE}$ are given as follows
\begin{eqnarray}\nonumber&&
{\rho}_{DE}=\frac{1}{2{\alpha}{\rho}-2}[3{\beta}{p}-({\beta}-12{\alpha}{H}^{2})
{\rho}-2{\alpha}{\rho}^{2}
+3{\alpha}{H}(\dot{\rho}-\dot{p})],
\\\label{12}&&{p}_{DE}=\frac{1}{2{\alpha}{\rho}-2}[-{\rho}({\beta}+6{\alpha}{H}^{2}+4{\alpha}{\dot{H}})
+p(-5{\beta}+12{\alpha}{H}^{2}-2{\alpha}{\rho}+4{\alpha}{\dot{H}})-
{\alpha}(4{H}({\dot{p}}+{\dot{\rho}})-{\ddot{p}}+{\ddot{\rho}})],
\end{eqnarray}
and effective EoS $\omega_{eff}$ is
\begin{eqnarray}
\omega_{eff}=\frac{-\rho(\beta+6\alpha{H}^2+4\alpha\dot{H})+p(-2
-5\beta+4\alpha(3{H}^2+\dot{H}))-\alpha(4H(\dot{p}+\dot{\rho})
-\ddot{p}+\ddot{\rho})}{3\beta{p}-(2+\beta-12\alpha{H}^2)\rho+3\alpha{H}
(\dot{\rho}-\dot{p})}.
\end{eqnarray}
The EoS of DE is, $\omega_{DE}=\frac{p_{DE}}{\rho_{DE}}$
\begin{eqnarray}
\omega_{DE}=\frac{-\rho(\beta+6\alpha{H}^2+4\alpha\dot{H})
+p(-5\beta+12\alpha{H}^2-2\alpha\rho+4\alpha\dot{H})
-\alpha(4H(\dot{p}+\dot{\rho})-\ddot{p}+\ddot{\rho})}
{3\beta{p}-(\beta-12\alpha{H}^2)\rho-2\alpha{\rho}^2+
3\alpha{H}(\dot{\rho}-\dot{p})},
\end{eqnarray}
and conservation equation (\ref{cons}) takes the form
\begin{eqnarray}\label{14}
{\dot{\rho}}+3{H}(p+\rho)=\frac{-18{\alpha}{H}^{3}(p+\rho)-6{\alpha}H(p+\rho){\dot{H}}
+3(\beta-\alpha\dot{H}){\dot{p}}+(\beta-3\alpha\dot{H})\dot{\rho}-9\alpha{H}^{2}
({\dot{p}}+{\dot{\rho}})}{2(1+\beta-6\alpha{H}^{2}-3{\alpha}{\dot{H}})}.
\end{eqnarray}
Now the above equations are expressed in terms of redshift by using relation $a(t)=\frac{1}{1+z}$, where ${\frac{d}{dt}}=\\-(1+z){H}{\frac{d}{dz}}$ whereas $p=p(z)$
and $\rho=\rho(z)$. Where prime is for derivative with respect to redshift parameter $z$.
\begin{eqnarray}\label{15}
&&\nonumber 3{H}(p+\rho)-(1+z){H}{\rho}^{\prime}=\frac{1}{2(1+\beta+3\alpha{H}
  (-2H+(1+z){H}^{\prime}))}\left[-(H(1+z)\beta({3}{p}^{\prime}+{\rho}^{\prime})
+9\alpha{H}^{2}(2{p}+2{\rho}-(1+z)\right.\\&&
\left.({p}^{\prime}+{\rho}^{\prime}))
+3(1+z)\alpha{H}{H}^{\prime}(-2{p}-2{\rho}+(1+z)({p}^{\prime}+
{\rho}^{\prime})))\right].
\end{eqnarray}
The revolutionary field equation $G_{\mu\nu}=8\pi G{T}_{\mu\nu}$ shows the connectedness of matter content of universe with geometry of the fabric of space-time, represented in Einstein's general theory of relativity. The LHS of the previously stated field equation show the Einstein tensor, which satisfy the Bianchi identities $\nabla_{\nu}G^\nu_\mu\equiv0$ and RHS shows the EMT.
If the covariance derivative of EMT is zero ($\nabla_\mu{T}^\nu_\mu=0$) then it shows the conservation of matter in every part of the universe. EFE can be explored on different choices of metric $g_{\mu\nu}$ and EMT $T_{\mu\nu}$. Although matter and geometry are on same footing but GR does not allow us to check the possible effects of  nonminimal coupling between them. These limitations of GR vanished in recently developed theories like $f(R,T)$ and $f(R,T,Q)$ theories. In these theories EMT is not conserved ($\nabla_\mu{T}^\nu_\mu\neq0$), we use this result to find the value of energy density. Such formation of energy density from the nonconserved EMT helps to study the role of non-minimal coupling in cosmic expansion. Before finding the value of $\rho(z)$ we should know the relation of $H(z)$. A lot of relations exists in literature with the requirement of their theoretical consistency and observational viability. But here we will take the power law expansion in terms of red shift given as $H(z)={H_0(1+z)}^{\frac{1}{m}}$, where $m$ is the power law exponent.

Power law cosmology appears as a good phenomenological explanation of the cosmic evolution, it can describe the cosmic history including radiation epoch, the dark matter epoch and the accelerating DE dominated epoch. Further, these solutions provide the scale factor evolution for the standard fluids such as dust matter case ($m=2/3$) or radiation dominated eras ($m=1/2$). Also, ${m\gtrsim1}$ predicts a late-time accelerating Universe. It provides an interesting alternative to deal with the problems like (age, flatness and horizon problems) associated with the standard model. Evolution of power law model has been discussed in various articles \cite{power}, for instance it addresses the horizon, flatness and age problems for the parametric value $m\geq1$ \cite{power1}. These type of solutions are found to be consistent with various data sets including nucleosynthesis \cite{kapl,Loh}, with the age of high-redshift objects such as globular clusters \cite{kapl,Loh}, with the SNIa data \cite{Sethidev}, and with X-ray gas mass fraction measurements of galaxy clusters \cite{AllenZhu}. In the framework of power law cosmology, authors have discussed the angular size-redshift data of compact radio sources \cite{Alcaniz}, the gravitational lensing statistics and SNIa magnitude-redshift relation \cite{Loh, Dev}.

\begin{figure}
\epsfig{file=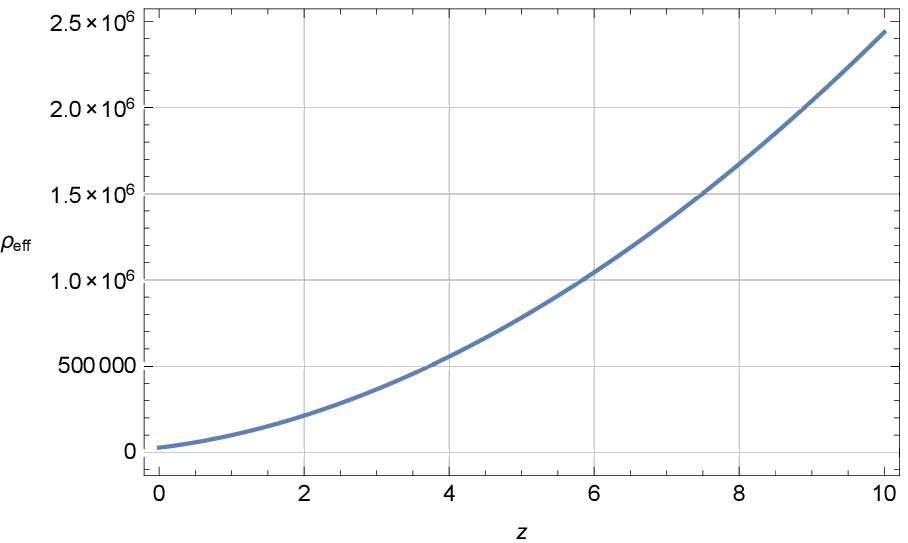,width=0.5\linewidth}\epsfig{file=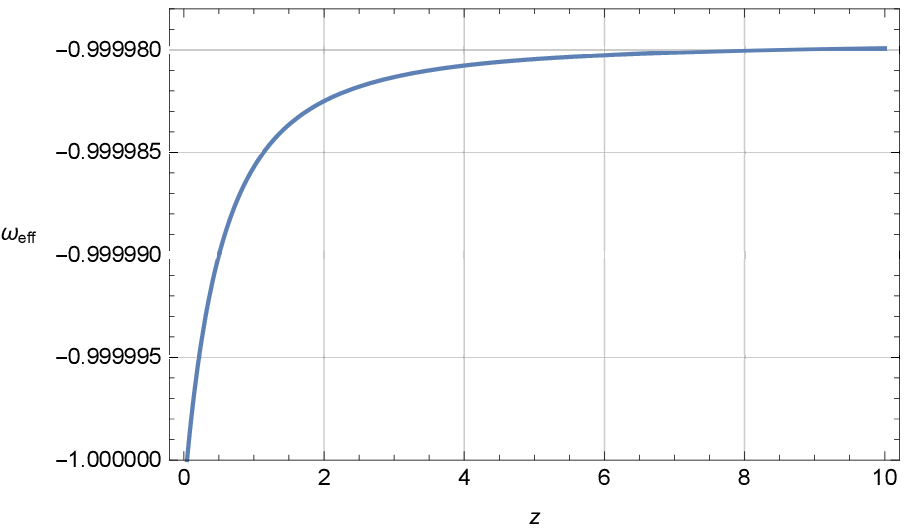,width=0.5\linewidth}
  \caption{Figure on the left represents the evolution of $\rho_{eff}$
  whereas the figure on the right side shows the behavior $\omega_{eff}$.
  Herein, we set $\alpha=10$, $\beta=-5$, $m=1.066658$, and $H_0=67.3$ \cite{2}.}
  \label{fig1}
\end{figure}

In this scenario, energy density is found by solving the Eq.(\ref{15}) as
\begin{eqnarray}\label{rho}
\rho(z)=e^{-\frac{(1+\omega)(6(1+\beta)\log(1+z)+\frac{m(-1+3m+\beta)
  (1+3\omega)\log[m(2+\beta-3\beta\omega)+3H_0^2(1+z)^{\frac{2}{m}}\alpha(1-\omega+m(-1+3\omega))]}
  {-1+m+\omega-3m\omega})}{-2+\beta(-1+3\omega)}}c,
\end{eqnarray}
where c is constant of integration. As energy density is found to be in an exponential form so it will remain positive for all values of unknowns parameters like $\alpha$, $\beta$, $\omega$, $m$, $z$. It will only depend on constant of integration c when we take negative value of c then energy density will be negative or less than zero otherwise for all positive values of c energy density will remain positive.
One can also get the relation between time and redshift as
\begin{equation}
t=\left(\frac{1}{1+z}\right)^{\frac{1}{m}}.
\end{equation}
Using the value of $\rho(z)$, one can get $\rho_{eff}$ and $p_{eff}$ in terms of redshift as and we take $c=10$ and $\omega=1$
\begin{equation}
\rho_{eff}=\frac{10(6H_0^2m(1+z)^{\frac{2}{m}}\alpha+m(2-2\beta))^{\frac{4(-1+3m+\beta)}{-2+2\beta}}
(-1+6 H_0^2(1+z)^{\frac{2}{m}}\alpha+\beta)}{-(1+z)^{\frac{12(1+\beta)}
{-2+2\beta}}+10\alpha(6H_0^2m(1+z)^{\frac{2}{m}}\alpha+m(2-2\beta))^{\frac{4(-1+3m+\beta)}{-2+2\beta}}},
\end{equation}
\begin{eqnarray}\nonumber&&
p_{eff}=\frac{1}{(1+z)^{\frac{6(1+\beta)}{-1+\beta}}-10\alpha(6H_0^2m(1+z)^{\frac{2}{m}}\alpha+m(2-2\beta))^{\frac{4(-1+3m+\beta)}
{-2+2\beta}}}2^{3+\frac{6m}{-1+\beta}}5(m(1+3H_0^2(1+z)^{\frac{2}{m}}\alpha-\beta))^{1+\frac{6m}{-1+\beta}}\\\label{12}&&
(3H_0^4(-16+21m)(1+z)^{\frac{4}{m}}\alpha^2-12H_0^2m(1+z)^{\frac{2}{m}}\alpha(2+\beta)-m(-1+\beta)(1+3\beta)),
\end{eqnarray}
and effective EoS in term of redshift can be written as
\begin{equation}
\omega_{eff}=\frac{48 H_0^4(1+z)^{\frac{4}{m}}\alpha^2+m(-1-63 H_0^4
(1+z)^{\frac{4}{m}}\alpha^2-2\beta+3\beta^2+12H_0^2(1+z)^{\frac{2}{m}}
\alpha(2+\beta))}{m(1+3H_0^2(1+z)^{\frac{2}{m}}\alpha-\beta)(-1+6H_0^2(1+z)^
{\frac{2}{m}}\alpha+\beta)}.
\end{equation}
Cosmic acceleration can be measured through a dimensionless cosmological function
known as the deceleration parameter $q$. Here, $q$ is given by
\begin{equation}
q=-\frac{a\ddot{a}}{{\dot{a}}^2}=\frac{1}{m}-1
\end{equation}
$q$ characterizes the accelerating or decelerating behavior of cosmos, here, $q < 0$ explains an accelerating epoch, whereas $q>0$ describes decelerating epoch. In power law cosmology we require $m>0$ to restrict $q$ as $q>-1$. Graphical representation of effective components $\rho_{eff}$, EoS $\omega_{eff}$ are shown in Fig.~\ref{fig1}. In this discussion, we choose the following values of unknown parameters $\alpha=10$, $\beta=-5$, and $m=1.066658$. For this value of $m$, deceleration parameter is $-0.0624924$ which favors the expanding behavior of cosmos. We set the parameters in a way to keep the positivity of $\rho_{eff}$. It can be seen that $\rho_{eff}$ is positive and increasing function as shown on right plot and $\omega_{eff}$ approaches to $-1$ at $z=0$ representing the $\Lambda$CDM epoch in accordance with recent observations from Plank's data \cite{2}.

\section{Energy conditions}

The EFE $G_{\mu\nu}=8\pi G{T}_{\mu\nu}$ describe the relation between space-time geometry and matter content. The LHS of this equation represent geometry and RHS corresponds the matter distribution. The suppressed idea is that the energy matter distribution tells us that how space time is curved and how gravity plays his role. Therefore, if we apply any condition on $T_{\mu\nu}$ then it will be immediately referred to the conditions on Einstein Tensor $G_{\mu\nu}$ \cite{36}. Matter energy distribution is responsible for casual and geodesic structure of space-time. For this purpose energy conditions ensure that the casuality principle is appreciated and acceptable physical sources have to be studied \cite{36,37}.
The energy conditions are based on Raychaudhuri equations and can be taken from the expansion given by
\begin{equation}
\frac{d\theta}{d\tau}=-\frac{\theta^2}{2}-\sigma_{\mu\nu}\sigma^{\mu\nu}+\omega_{\mu\nu}\omega^{\mu\nu}-R_{\mu\nu}k^\mu k^\nu
\end{equation}
where $\theta$, $\sigma_{\mu\nu}$ and $\omega_{\mu\nu}$  shows expansion, shear and rotation respectively. These parameters are related to the congruence explained by the null vector field $k^\mu$.
The shear is a spatial tensor with $\sigma^2\equiv\sigma_{\mu\nu}\sigma^{\mu\nu}\geq0$, thus it is obvious from Raychaudhuri equations that for any hypersurface orthogonal congruences, which forces $\omega\equiv0$, the condition for attractive gravity reduce to $R_{\mu\nu}k^\mu k^\nu\geq0$. However, in GR, through the EFE we can write $T_{\mu\nu}k^\mu k^\nu\geq0$.
In the context of modified theory we used the effective EMT which is shown in equation (\ref{5}) and positivity condition, $R_{\mu\nu}k^\mu k^\nu\geq0$ in the Raychaudhuri equations gives the following form of NEC $T_{\mu\nu}^{eff}k^\mu k^\nu\geq0$ and for ordinary matter we can also write $T_{\mu\nu}^{mat}k^\mu k^\nu\geq0$. It is simple to prove that the previous conditions impose energy density positive in all local frame of references by using local lorentz transformation. Energy conditions  describes the behavior of the similarity of lightlike, timelike and spacelike curves. It is generally used in GR to find and study the singularities of space time.
The energy conditions, NEC (Null energy condition), WEC (Weak energy condition), SEC (Strong energy condition) and DEC (Dominant energy condition) in terms of EMT are given by
\begin{eqnarray}\nonumber
NEC\Leftrightarrow \rho_{eff} + p_{eff}\geq0,\\\nonumber
WEC\Leftrightarrow \rho_{eff} \geq0 ,~~ \rho_{eff} + p_{eff} \geq0,\\\nonumber
SEC\Leftrightarrow \rho_{eff}+3p_{eff}\geq0,~~  \rho_{eff}+p_{eff}\geq0,\\\nonumber
DEC\Leftrightarrow \rho_{eff}\geq0  ,~~  \rho_{eff}\pm p_{eff}\geq0.\\\nonumber
\end{eqnarray}
Now, we will discus the energy conditions for our particular model of $f(R,T,Q)$ gravity which is $R+\alpha Q+\beta T$ by considering FLRW metric.
WEC is found to be of the following form
\begin{eqnarray}\label{wec}&&
\textbf{WEC}: {\rho}_{eff}=\rho+\frac{1}{2{\alpha}{\rho}-2}[3{\beta}{p}-({\beta}-12{\alpha}{H}^{2})
{\rho}-2{\alpha}{\rho}^{2}
+3{\alpha}{H}(\rho^\prime-p^\prime)]\geq0,
\end{eqnarray}
NEC yields as
\begin{eqnarray}\nonumber&&
\textbf{NEC}: \rho_{eff}+p_{eff}=\frac{1}{2\alpha\rho-2}[-2p(1+\beta-6\alpha
H^2+2(1+z)\alpha H H^\prime)+\rho(-2(1+\beta)+6\alpha H^2+4(1+z)
\alpha H H^\prime)\\\label{nec} &&+(1+z)\alpha H((1+z)H^\prime(p^\prime-\rho^\prime)+H(8p^\prime+(1+z)
(p^{\prime\prime}-\rho^{\prime\prime})))]\geq0.
\end{eqnarray}
SEC yields as
\begin{eqnarray}\nonumber&&
\textbf{SEC}: \rho_{eff}+3p_{eff}=\frac{1}{2\alpha\rho-2}[-2\rho(1+2\beta+3\alpha
H^2-6(1+z)\alpha H{H}^\prime)-6p(1+2\beta-6\alpha H^2+2(1+z)
\alpha H H^\prime)\\\label{sec}&&+3(1+z)\alpha H((1+z)H^\prime(p^\prime-\rho^\prime)+H(6p^\prime+2\rho^\prime+(1+z)
(p^{\prime\prime}-\rho^{\prime\prime})))]\geq0.
\end{eqnarray}
DEC yields as
\begin{eqnarray}\nonumber&&
\textbf{DEC}: \rho_{eff}-p_{eff}=\frac{1}{2\alpha\rho-2}[2\rho(-1+9\alpha
H^2-2(1+z)\alpha H{H}^\prime)+2p(1+4\beta-6\alpha H^2+2(1+z)
\alpha H H^\prime)-\\\label{dec}&&(1+z)\alpha H((1+z)H^\prime(p^\prime-\rho^\prime)+H(2p^\prime+6\rho^\prime+(1+z)
(p^{\prime\prime}-\rho^{\prime\prime})))]\geq0.
\end{eqnarray}

The inequalities $(\ref{wec}$-$\ref{dec})$ depends on five parameters $\alpha$, $\beta$, $m$, $z$, $c$. In this approach, we fixed two parameters and find the valid regions by varying the possible ranges of other parameters. We prefer to fix the constant of integration as $c=10$ and range of $z$ will be from $-0.9$ to $10$ and show the results for WEC and NEC. The validity region for different cases are shown in TABLE \textbf{1} in which we took the different values of $m$ to show the relation between $\alpha$, $\beta$ and $m$. Initially, we fix the value of $m=1.1$ for WEC then range for alpha is $\alpha\geq 1.01$ and for beta is
($1.1\leq\beta\leq270$) and ($-550\leq\beta\leq-2.8$). If we take value of $m=10$ then the ranges of $\beta$ will also increase like for $\alpha\geq 1.01$, it requires ($1.1\leq\beta\leq9000$) and ($-17900\leq\beta\leq-50$). We can see that WEC is valid only for positive values of $\alpha$ whereas $\beta$ needs some particular range for different values of $\alpha$ and $m$. If we take small value of $m$ then validity range is also small for $\beta$, likewise if we increase the starting value of $\alpha$ then range of $\beta$ also increases. Choice of $m$ and particular range of $\beta$ are directly proportional to each other while $\alpha\geq 1.01$ and $\alpha$ has also direct relation with $\beta$. If we will take larger value of $\alpha$ then we have to choose the larger value for $\beta$ and vice versa, like if we choose $m=2$ then $\alpha=10$ and $\beta=-5$ but if decrease the value of alpha as $\alpha=1.001$ then $\beta$ will be $-8$.

$\rho_{eff}+p_{eff}\geq0$ is also valid for positive values of $\alpha$. In this setup, we show different ranges of $\beta$ depending on the particular ranges of $\alpha$ and results are shown in TABLE \textbf{1}. If we choose $m=1.1$ with $\alpha\geq1.01$ then range of $\beta$ is ($-1.3\leq\beta\leq0.9$). If we fix $m=2$ and $\alpha\geq1.01$ then range of $\beta$ is ($-2.9\leq\beta\leq0.9$). From this discussion we can conclude that range of $\beta$ for $\rho_{eff}+p_{eff}\geq0$ lies between $-2.9$ and $0.9$ for any value of $\alpha\geq1.01$ and $m>1$. Finally, in the last two columns of TABLE \textbf{1} we show the combine validity region for WEC and NEC. Same in this case if we increase the value of $m$ then the ranges of $\alpha$ also increases. Keep in notice that in common region, range of $\alpha$ is also restricted and very short. For $m=1.1$ range of $\alpha$ is ($0\leq\alpha\leq0.00000016$) and range of $\beta$ is ($-0.64\leq\beta\leq0.11$). If we fix $m=10$ then range of $\alpha$ is ($0\leq\alpha\leq0.00001$) and range of $\beta$ is ($-0.55\leq\beta\leq0.22$) for $\rho_{eff}$.

\begin{figure}
  \epsfig{file=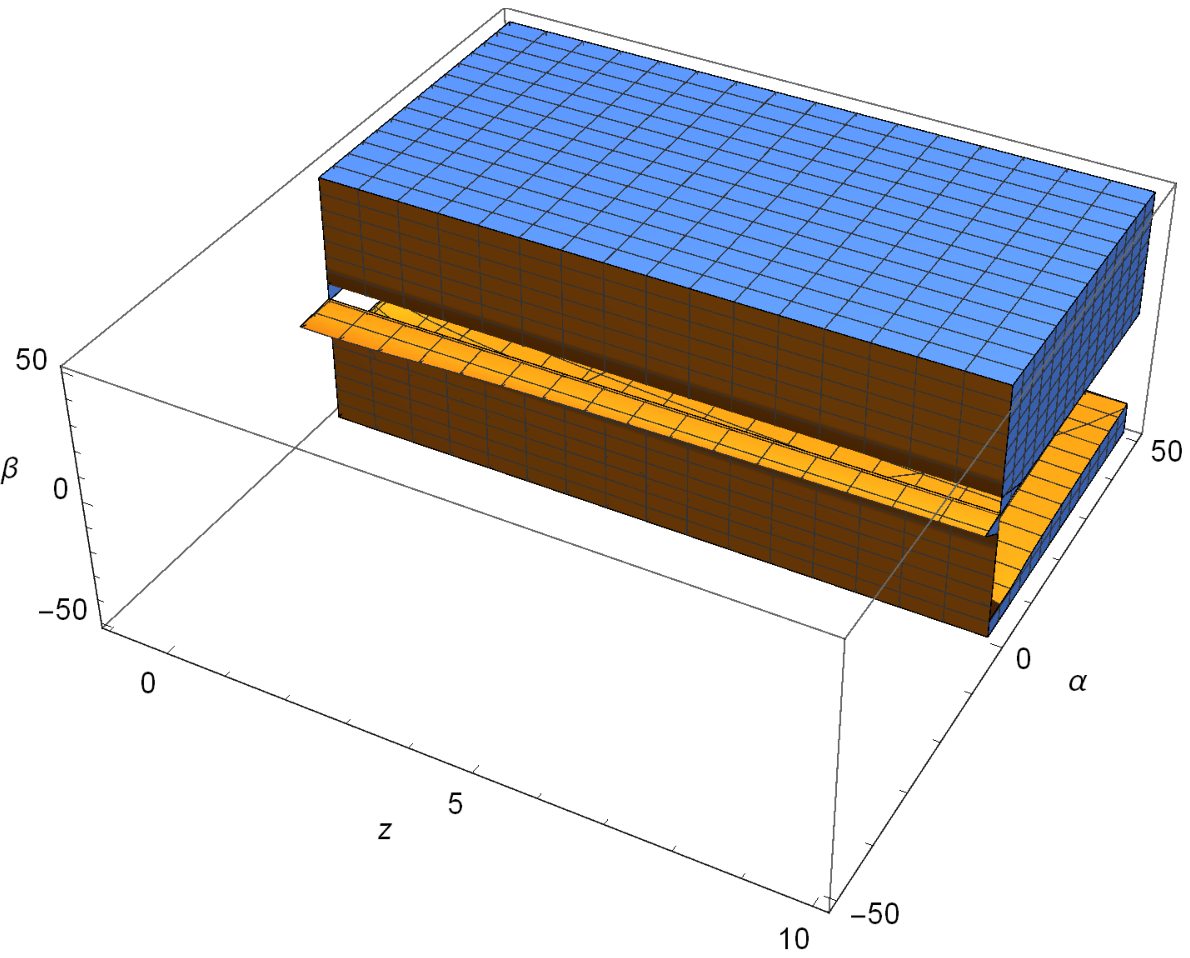,width=0.5\linewidth}\epsfig{file=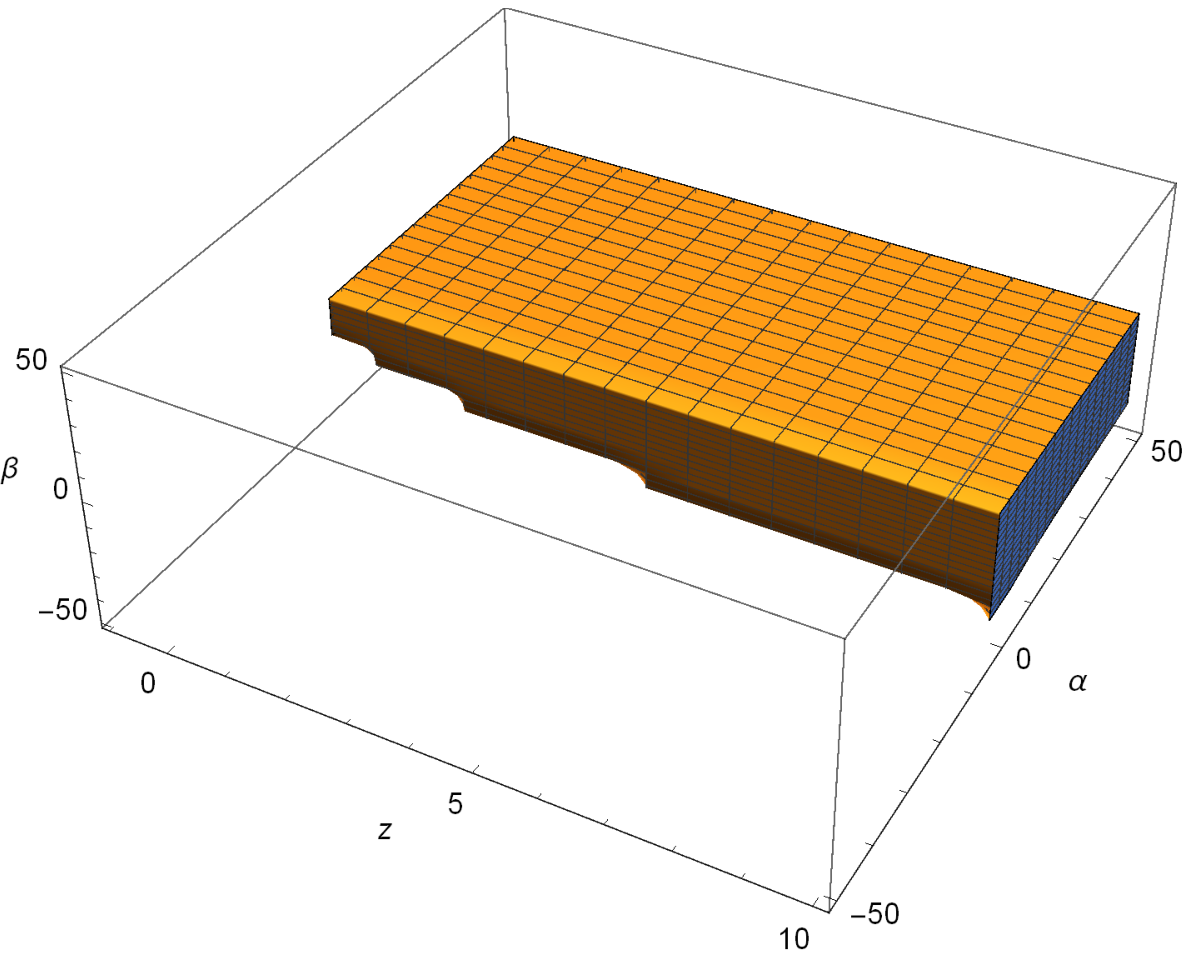,width=0.5\linewidth}
  \caption{Figure on the left represents the validity region for $\rho_{eff}\geq0$
  whereas the figure on the right side shows the validity region for $\rho_{eff}+p_{eff}\geq0$.
  Herein, we set $H_0=67.3$ versus $z$.}
  \label{fig2}
\end{figure}

\begin{figure}
  \epsfig{file=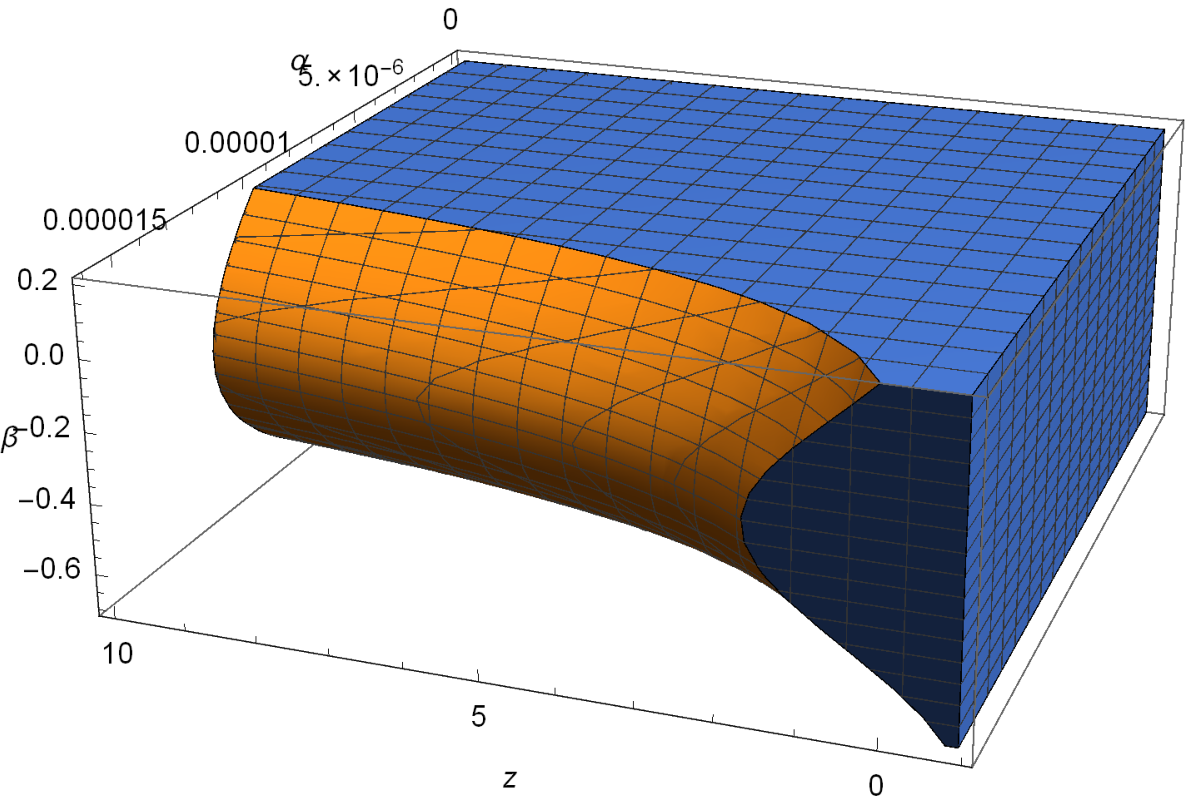,width=0.5\linewidth}\epsfig{file=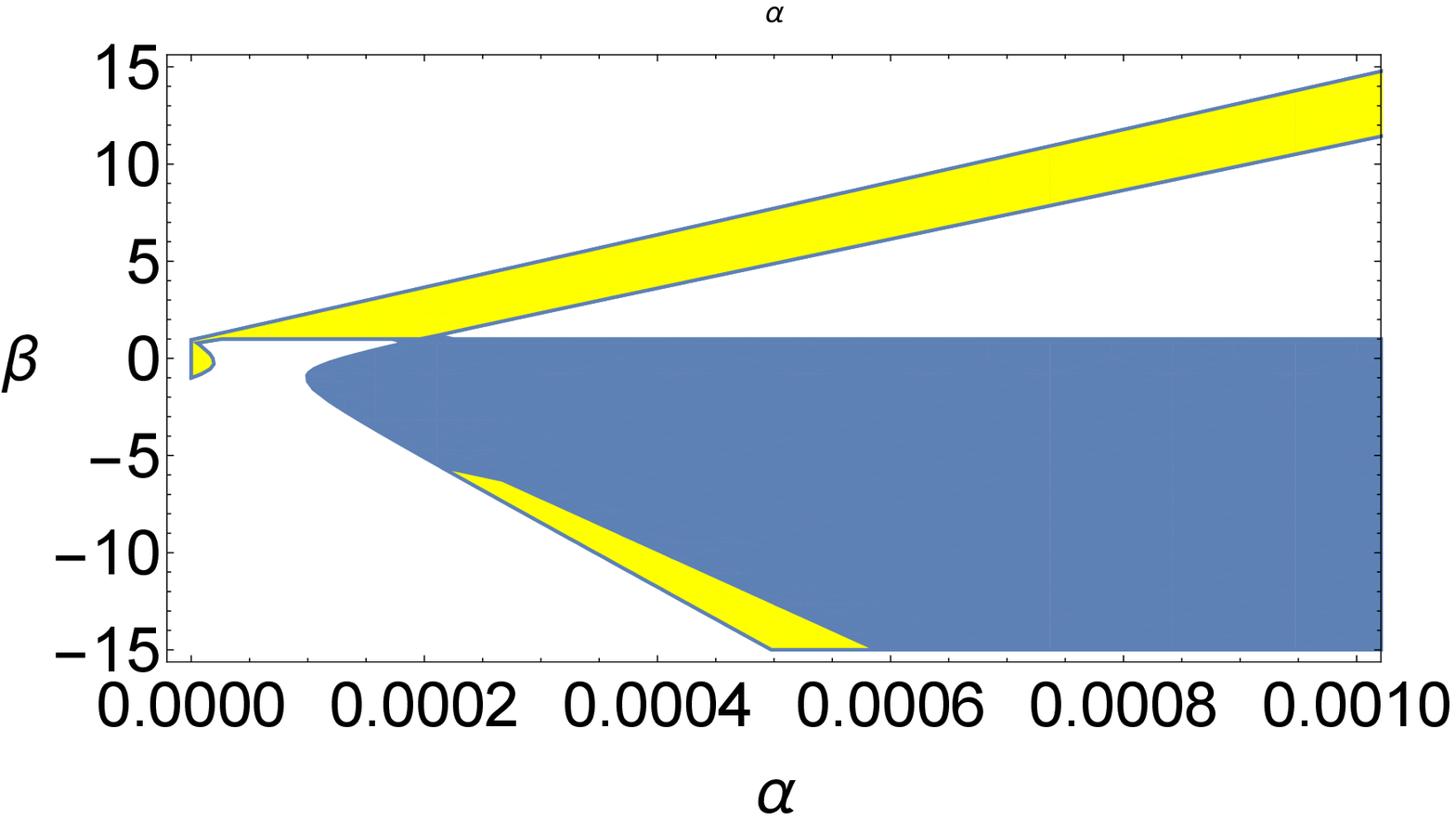,width=0.5\linewidth}
  \caption{Figure on the left represents the validity region for WEC ($\rho_{eff}\geq0$, $\rho_{eff}+p_{eff}\geq0$) in $3D$  whereas the figure on the right side also shows the validity region for WEC in $2D$.
  Herein, we set $H_0=67.3$ versus $z$, and for $2D$ plot we set $z=0$.}
  \label{fig3}
\end{figure}

We show the graphical description for validity regions of $\rho_{eff}\geq 0$ and $\rho_{eff}+p_{eff}\geq 0$ in Figs.~\ref{fig2}-\ref{fig3}. In Fig.~\ref{fig2}, we show the validity region for $\rho_{eff}\geq 0$ $\rho_{eff}+p_{eff}\geq 0$ for the particular choice $m=10$. Fig.~\ref{fig3}, shows the region which validate the NEC for $m=10$. Right side of Fig.~\ref{fig3} presents the common region for both $\rho_{eff}\geq 0$  and $\rho_{eff}+p_{eff}\geq 0$ at $z=0$ and $m=10$. Yellow color shows the region of $\rho_{eff}\geq0$ and blue color shows the region for $\rho_{eff}+p_{eff}\geq0$.
The validity regions of energy conditions are shown in TABLE \textbf{1}.
\begin{table}[H]
  \centering
  \begin{tabular}{|l|c|c|c|c|c|c|c|}\hline
  \multicolumn{3}{|c|}{$\rho_{eff}\geq0$}&\multicolumn{2}{|c|}{$\rho_{eff}+p_{eff}\geq0$}&
  \multicolumn{2}{|c|}{$\rho_{eff}+p_{eff}\geq0$, $\rho_{eff}\geq0$}  \\ \hline
m & Validity of $\alpha$& Validity of $\beta$&
Validity of $\alpha$& Validity of $\beta$&Validity of $\alpha$& Validity of $\beta$ \\  \hline \hline
1.1 & &$1.1\leq\beta\leq270$
& &$-1.3\leq\beta\leq0.9$&
 $0\leq\alpha\leq0.00000016$&$-0.64\leq\beta\leq0.11$\\
 &  & $-550\leq\beta\leq-2.8$ &
 & &
  & \\ \cline{1-1} \cline{3-3} \cline{5-7}
2 & $\alpha\geq1.01$ & $1.1\leq\beta\leq1500$ &
 $\alpha\geq1.01$ & $-2.9\leq\beta\leq0.9$&
 $0\leq\alpha\leq0.0000013$ & $-0.58\leq\beta\leq0.29$\\
 &  & $-3000\leq\beta\leq-8$ &
  & &
  & \\ \cline{1-1} \cline{3-3} \cline{5-7}
10 &  & $1.1\leq\beta\leq9000$
& & $-17.8\leq\beta\leq0.9$&
 $0\leq\alpha\leq0.00001$ & $-0.55\leq\beta\leq0.22$\\
 &  & $-17900\leq\beta\leq-50$ &
  & &
  & \\
\hline
\end{tabular}
\caption{Validity ranges of parameters $\alpha$ and $\beta$.}\label{1}
\end{table}
One can represent energy conditions in the combined form as
\begin{equation}
\beta A_1+\alpha H A_2\geq A_3
\end{equation}
where $A_{i,s}$ purely depend on energy conditions which are under discussion
for WEC, we found the values of $A_{i,s}$
\begin{eqnarray}\nonumber
&&{A}_1^{WEC}=\frac{3p-\rho}{2},  A_2^{WEC}=6\rho H +\frac{3}{2}(1+z)H(p^\prime-\rho^\prime)\\&& A_3^{WEC}=\rho
\end{eqnarray}
for NEC, we can found as
\begin{eqnarray}\nonumber
&&{A}_1^{NEC}=-\rho-p,     A_2^{NEC}=p(6H-2(1+z)H^\prime)+\rho(3H+2(1+z)H^\prime)+\frac{1+z}{2}[(1+z)H^\prime(p^\prime-\rho^\prime)
+H(8p^\prime+\\&&(1+z)(p^{\prime\prime}-\rho^{\prime\prime}))], A_3^{NEC}=p+\rho
\end{eqnarray}
for SEC, we can found as
\begin{eqnarray}\nonumber
&&{A}_1^{SEC}=-2\rho-6p,     A_2^{SEC}=3p(6H-2(1+z)H^\prime)+\rho(-3H+6(1+z)H^\prime)+\frac{3(1+z)}{2}[(1+z)H^\prime(p^\prime-\rho^\prime)
+H(6p^\prime+\\&&2\rho^\prime+
(1+z)(p^{\prime\prime}-\rho^{\prime\prime}))], A_3^{SEC}=\rho+3p
\end{eqnarray}
for DEC, we can found as
\begin{eqnarray}\nonumber
&&{A}_1^{DEC}=4p,     A_2^{DEC}=p(-6H+2(1+z)H^\prime)+\rho(9H-2(1+z)H^\prime)-\frac{1+z}{2}[(1+z)H^\prime(p^\prime-\rho^\prime)
+H(2p^\prime+6\rho^\prime+\\&&(1+z)(p^{\prime\prime}-\rho^{\prime\prime}))], A_3^{DEC}=\rho-p
\end{eqnarray}
\section{Discussion}

Modified theories become the appropriate candidates to discus the accelerated cosmic expansion. $f(R,T,Q)$ is a generalized modified theory based on curvature matter coupling. This complicated theory involve the contraction of EMT and Ricci tensor $R_{\mu\nu}$ and it is the extended form of $f(R,T)$ gravity \cite{22}. Although, $f(R,T,Q)$ gravity is the extension of $f(R,T)$ theory but there exists a notable difference while taking the contraction term $R_{\mu\nu}$. For instance, if we involve the role of electromagnetic field  or radiation dominated fluid, then filed equations turn down to $f(R)$ gravity and effect of non-minimal coupling would be vanished in $f(R,T)$ gravity but this is not the case for $f(R,T,Q)$ gravity due to the involvement of contraction term $Q$. $Q$ is the inclusive term which involve the strong non-minimal coupling as compared to other modified theories. Indeed, in this theory the interaction between matter and geometry can be seen through the coupling of energy-momentum and Ricci tensors. $Q$ is the generic term responsible for the non-minimal coupling as compared to other modified theories. In fact the fundamental characteristic of theories involving non-minimally matter geometry coupling is the non-conserved EMT produced from the divergence of the field equations. As a result, motion of test particles is non-geodesic and an extra force orthogonal to four-velocity of the particle is present due to matter geometry coupling. This is consistent with the interpretation of four force which states that component of force orthogonal to particle's four-velocity can influence its trajectory. It also involves the contribution from the Ricci tensor and may lead to significant deviation from the geodesic
paths. The extra force can be useful to explain the dark matter properties and Pioneer
anomaly. One can count the additional curvature obtained from the curvature matter
coupling to inform the galactic rotation curves. This theory can also present novel views
about the early stages of cosmic evolution specifically the inflationary paradigm. The
non-minimal theories as that of this theory imply the violation of equivalence principle.
Similar behavior is also suggested in cosmological study, e.g., Bertolami et al. \cite{Ber07}
showed that data from Abell cluster A586 supports the interaction between dark matter
and energy which does imply the violation of equivalence principle. Thus it would be
interesting to test these models with non-minimal coupling and explore their implications
in cosmology, gravitational collapse as well as in gravitational waves.

In this article, we have constructed a cosmological scenario from the complicated non-minimal matter geometry coupling in the $f(R,T,Q)$ gravity. We consider a simplest case of non-minimal coupling in this modified theory in the form of model $f(R,T,Q)=R+\alpha T+\beta Q$. Dynamical equations are presented in section \textbf{III}, where we consider the power law cosmology to find an expression for energy density $\rho$. Using Eq.(\ref{rho}), it is obvious to find the expressions of effective enregy momentum tensor and its components. In power law cosmology, one can represent the cosmic history depending on the choice of parameter $m$. Here, we set parameter $m$ according to the evolution of $q$ as per recent observational data. In Fig.~\ref{1}, we set $m=1.0666580$ with $q=-0.0624924$ to see the evolution of $\rho_{eff}$ and $\omega_{eff}$, it is found that WEC is satisfied and $\omega_{eff}\rightarrow-1$ validating the current cosmic epoch \cite{2}. It is to be noted that we set the choice of parameters $\alpha$ and $\beta$ as per validity ranges expressed in Table \textbf{1}, where we develop the constraints on these parameters for different values of $m$ satisfying WEC and NEC. Evolution of WEC and NEC versus redshift $z$ is presented in Fig.~\ref{2}-\ref{4}.

In literature, observational constraints have been developed on the choice of power law exponent $m$, cosmological parameters $q$ and $H_0$. Kaeonikhom et al. \cite{keon} explored the phantom power law cosmology using cosmological observations from Cosmic Microwave Background (CMB), Baryon Acoustic Oscillations (BAO) and observational Hubble data, they found the best fit value of power law exponent as $m\approx-6.51^{+0.24}_{-0.25}$. In \cite{Kumar}, Kumar found the constraints on Hubble and deceleration parameters from the latest $H(z)$ and SNe Ia data as {$q=-0.18^{+0.12}_{-0.12}$, $H_0=68.43^{+2.84}_{-2.80}$kms-1Mpc-1} and {$q=-0.38^{+0.05}_{-0.05}$, $H_0=69.18^{+0.55}_{-0.54}$ kms-1Mpc-1} respectively. The combination of $H(z)$ and SNe Ia data yields the constraints {$q=-0.34^{+0.05}_{-0.05}$, $H_0=69.18^{+0.55}_{-0.54}$kms-1Mpc-1}. The consistent observational constraints on both of the parameters $q$ and $H_0$ according to latest $28$ points of $H(z)$ are found as {$q=-0.0451^{+0.0.0614}_{-0.0625}$, $H_0=65.2299^{+2.4862}_{-2.4607}$, in case of Union2.1 SN data, these parameters take the values {$q=-0.3077^{+0.1045}_{-0.1036}$, $H_0=68.7702^{+1.4052}_{-1.3754}$ \cite{rani}. Using the data set of Kumar \cite{Kumar} and Rani et al. \cite{rani}, we choose the parameter $m$ and develop the ranges of $\omega_{eff}$ as shown in Table \textbf{2}. For $m=1.221$, $\omega_{eff}$ is found to be $-1.31601$ which agrees with the observational results of Planck$+$WMAP$+H_0$ \cite{2}. Also, for the choice of $m=1.0473$ and $m=1.4445$, results of $\omega_{eff}$ are consistent with the observational data of $95\%$(WMAP5$+$BAO$+$SN) \cite{komatsu} and WMAP9 \cite{hinshaw}.

\begin{table}[H]
  \centering
  \begin{tabular}{|l|c|c|c|c|c}\hline
  Data               &           $q$                   &           $H_0(km s^{-1}Mpc^{-1})$     &   $m$     & $\omega_{eff}$  \\\hline
  $H(z)$ (14 points) \cite{Kumar}&  $-0.18^{+0.12}_{-0.12}$       & $68.43^{+2.84}_{-2.80}$       & 1.221  & $-1.31601$       \\\hline
  SN (Union2) \cite{Kumar}        &  $-0.38^{+0.05}_{-0.05}$       & $69.18^{+0.55}_{-0.54}$       & 1.613  & $-1.84677$
  \\\hline
  $H(z)+SN (Union2)$ \cite{Kumar}&  $-0.34^{+0.05}_{-0.05}$       & $68.93^{+0.53}_{-0.52}$       & 1.516  & $-1.74099$
  \\\hline
  $H(z)$ (29 points) \cite{rani}&  $-0.0451^{+0.0614}_{-0.0625}$ & $65.2299^{+2.4862}_{-2.4607}$ & 1.0473 & $-0.953795$      \\\hline
  SN (Union2.1)\cite{rani}      &  $-0.3077^{+0.1045}_{-0.1036}$ & $68.7702^{+1.4052}_{-1.3754}$ & 1.4445 & $-1.65392$       \\\hline
\end{tabular}
\caption{Observational data results for power law exponent $m$ and $\omega_{eff}$.}\label{2}
\end{table}

\section*{Acknowledgments}

``M. Zubair thanks the Higher Education Commission, Islamabad, Pakistan for its
financial support under the NRPU project with grant number
$\text{5329/Federal/NRPU/R\&D/HEC/2016}$''.

\vspace{.5cm}


\begin{thebibliography}{36}
\bibitem{1}Perlmutter, S. et al., Astrophys. J. \textbf{517}, 565 (1999).
\bibitem{2}Ade, P.A.R., et al., Astronomy and Astrophysics. \textbf{571}, A1 (2014); Spergel, D. N., et al., Astrophys. J. Suppl. Ser. \textbf{170}, 37
(2007).
\bibitem{3}Cole, S. et al., Mon. Not. R. Astron. Soc. \textbf{362}, 505 (2005).
\bibitem{4}Hawkins, E. et al., Mon. Not. R. Astron. Soc. \textbf{346}, 78 (2003); Tegmark, M., et al., Phys. Rev. \textbf{D 69}, 103501 (2004).
\bibitem{5}Jain, B., Taylor, A., Phys. Rev. Lett. \textbf{91}, 141302 (2003).
\bibitem{6}P.A.R. Ade et al. ,
  Astronomy and Astrophysics. \textbf{571}, A16 (2014).
\bibitem{7}Caldwell, R.R. 
Phys. Lett. \textbf{B 23}, 545 (2002).
\bibitem{8}Sahni, V., Starobinsky, A.A. 
Int. J. Mod. Phys. \textbf{D 9}, 373 (2000).
\bibitem{9}Akarsu, O., Kilinc, B.C. 
Gen. Relativ. Gravitation. \textbf{42}, 119 (2010).
\bibitem{10}Nojiri, S. and Odintsov, S.D. 
Int. J. Geom. Meth. Mod. Phys. \textbf{4}, 115 (2007); Sotiriou, T.P. and Faraoni, V. 
Rev. Mod. Phys. \textbf{82}, 451 (2010); Nojiri, S. and Odintsov, S.D. 
Phys. Rept. \textbf{505}, 59 (2011);
    Sharif, M. and Zubair, M. 
     Astrophys. Space Sci. \textbf{342}, 511 (2012); Bamba, K. Capozziello, S. Nojiri, S. and Odintsov, S.D. 
     Astrophys. Space Sci. \textbf{345}, 155 (2012); Sharif, M. and Zubair, M. 
    \textbf{2013}, 790967 (2013).
\bibitem{11}Brans, C. and Dicke, R. 
 phys. Rev. \textbf{124}, 925 (1961).
\bibitem{12}Allemandi, G. Borowiec, A. Francaviglia, M. and Odintsov, S.D. 
 Phys. Rev.  \textbf{D 72}, 063505 (2005); Inagaki, T. Nojiri, S. and
Odintsov, S.D.  
JCAP. \textbf{06}, 010 (2005).
\bibitem{13}Bertolami, O. Boehmer, C.G. Harko, T. and Lobo, F.S.N.  
 Phys. Rev. \textbf{D 75}, 104016 (2007).
\bibitem{14}Bertolami, O. and Paramos, J.  
Class. Quantum Grav. \textbf{25}, 245017 (2008).
\bibitem{15}Bertolami, O. Lobo, F.S.N. and Paramos, J. 
Phys. Rev. \textbf{D 78}, 064036 (2008).
\bibitem{16}Harko, T. 
Phys. Lett.  \textbf{B 669}, 376 (2008).
\bibitem{17}Wu, Y.-B. 
Phys. Lett. \textbf{B 717}, 323 (2012).
\bibitem{18}Harko, T. and Lobo, F.S.N.  
Eur. Phys. J. \textbf{C 70}, 373 (2010).

\bibitem{zfRLM}Sharif, M. and Zubair, M. Adv. High Energy Phys. \textbf{2013} 947898 (2013).
\bibitem{baha} S.~Bahamonde, M.~Marciu and P.~Rudra,
  JCAP {\bf 1804}, no. 04, 056 (2018); S.~Bahamonde,
  Eur.\ Phys.\ J.\ C {\bf 78}, no. 4, 326 (2018); S.~Bahamonde, M.~Zubair and G.~Abbas,
  Phys.\ Dark Univ.\  {\bf 19}, 78 (2018). 


\bibitem{19}Sotiriou, T.P. and Faraoni, V. 
Class. Quantum Grav. \textbf{25}, 205002 (2008).
\bibitem{20}Bertolami, O. and Paramos, J. 
Class. Quantum Grav. \textbf{25}, 245017 (2008).
\bibitem{21}Poplawski, N.J. 
arXiv:gr-qc/0608031.
\bibitem{22}Harko, T. Lobo, F.S.N. Nojiri, S. and Odintsov, S.D.  
 Phys.Rev. \textbf{D 84}, 024020 (2011).
\bibitem{23}Sharif, M. and Zubair, M.  
 JCAP. \textbf{03}, 028 (2012); Sharif, M. and Zubair, M.  
 J. Exp. Theor. Phys. \textbf{117}, 248 (2013).
\bibitem{24}Houndjo, M.J.S. and Piattella, O.F. 
Int. J. Mod. Phys. \textbf{D 21}, 1250024 (2012); Sharif, M. and Zubair, M. 
J. Phys. Soc. Jpn. \textbf{82}, 064001 (2013).
\bibitem{25}Houndjo, M.J.S. 
Int. J. Mod. Phys. D \textbf{21}, 1250003 (2012).
\bibitem{26}Sharif, M. and Zubair, M. 
J. Phys. Soc. Jpn. \textbf{81}, 114005 (2012).
\bibitem{27}Jamil, M. Momeni1, D, Raza M. and Myrzakulov, R. 
Eur. Phys. J. \textbf{C 72}, 1999 (2012).
\bibitem{28}Alvarenga, F.G. et al.  
Phys. Rev. \textbf{D 87}, 103526 (2013).
\bibitem{29}Sharif, M. and Zubair, M.  
 Gen. Relativ. and Gravitation. \textbf{46}, 1723 (2014).
\bibitem{30}Shabani, H. and Farhoudi, M. 
 Phys. Rev. \textbf{D 88}, 044048 (2013).
\bibitem{fRTlit}Moraes, P.H.R.S. Correa, R.A.C. Lobato, R.V. JCAP \textbf{07}, 029 (2017); Moraes, P.H.R.S. Sahoo, P.K. Phys. Rev.  \textbf{D 96}, 044038 (2017); Noureen, I. et al., Eur. Phys. J.  \textbf{ 75}, 323 (2015); Noureen, I., Zubair, M. Eur. Phys. J. \textbf{C 75}, 62 (2015); Moraes, P.H.R.S.  Correaa,  R.A.C. and Lobato, R.V. JCAP\textbf{07}, 029 (2017); Shamir, M.F. Eur. Phys. J.  \textbf{C 75} 354 (2015); Zubair, M. Abbas, G. Noureen, I. Astrophys Space Sci  361:8 (2016); Zubair, M. Sardar, ·I.H. Rahaman, ·F. Abbas, ·G. Astrophys Space Sci  361:238 (2016); Zubair, M. Hina Azmat and Ifra Noureen, Eur. Phys. J.  \textbf{C 75} 62 (2015); Zubair, M. and Noureen, I.: Eur. Phys. J. C \textbf{75} 265 (2015); Hina Azmat, Zubair, M. and Ifra Noureen, Int. J. Mod. Phys. \textbf{D 27} 1750181 (2018); Zubair, M. Hina Azmat and Ifra Noureen, Int. J. Mod. Phys.  \textbf{D 27} 1850047 (2018); Zubair, M. Abbas, G. Noureen, I. Astrophys Space Sci  361:8 (2016); Zubair, M. Sardar, I.H. Rahaman, F. and Abbas, G. Astrophys Space Sci  361:238 (2016).

\bibitem{31}Odintsov, S.D. and Saez-Gomez, D.  
Phys. Lett. \textbf{B 725}, 437 (2013).
\bibitem{32}Haghani, Z. Harko, T. Lobo, F.S.N. Sepangi, H.R. and Shahidi, S. 
 Phys. Rev. \textbf{D 88}, 044023 (2013).
\bibitem{33}Sharif, M. and Zubair, M. 
 JCAP. \textbf{11}, 042 (2013).
\bibitem{34}Sharif, M. and Zubair, M. 
JHEP. \textbf{12}, 079 (2013).
\bibitem{35}Baffou, E.H., Houndjo, M.j.S. and Tosssa, J. 
 Astrophys. space. Sci. \textbf{361}, 376 (2016).
\bibitem{coupling} Koivisto, T. Classical Quantum Gravity \textbf{23}, 4289 (2006); Bertolami, O. Boehmer, C.  Harko, T. and  Lobo, F. S. N.
Phys. Rev. \textbf{D 75}, 104016 (2007);  Bertolami, O.  Paramos, J.  Harko, T. and  Lobo, F. S. N. arXiv:0811.2876.

\bibitem{power}Lohiya D. and  Sethi, M. Class. Quant. Grav. \textbf{16} 1545 (1999) [gr-qc/9803054] [INSPIRE].
 Sethi, M.  Batra A. and  Lohiya, D. Phys. Rev.  \textbf{D 60}  108301 (1999);  Batra, A.  Lohiya, D.  Mahajan S. and  Mukherjee, A. Int. J. Mod. Phys. \textbf{D 9}  757 (2000);  Gehlaut, S.  Geetanjali P.K. and  Lohiya, D. 
astro-ph/0306448 [INSPIRE];  Dev, A.  Jain D. and  Lohiya, D. 
arXiv:0804.3491;  Dev, A.  Safonova, M.  Jain D. and  Lohiya, D. Phys. Lett.  \textbf{B 548}  12 (2002) [astro-ph/0204150] [INSPIRE].
\bibitem{power1}Sethi, G. Dev, A. Jain, D.  Phys. Lett.  \textbf{B 624}, 135 (2005).

\bibitem{kapl}  Kaplinghat, M.  Steigman, G.  Tkachev, I.  Walker, T.P. Phys. Rev. \textbf{D 59} 043514 (1999).
\bibitem{Loh}  Lohiya, D.  Sethi, M. Class. Quantum Grav. \textbf{16} 1545 (1999).
\bibitem{Sethidev} Sethi, G.  Dev, A.  Jain, D. Phys. Lett. \textbf{B 624}  135 (2005);  Dev, A.  Jain, D.  Lohiya, D. arXiv:0804.3491 [astro-ph].
\bibitem{AllenZhu} Allen, S.W.  Schmidt, R.W.  Ebeling, H.  Fabian, A.C.  van  Speybroeck, L. Mon. Not. Roy. Astron. Soc. \textbf{353}  457 (2004); Zhu, Z.H.  Hu, M.  Alcaniz, J.S.  Liu, Y.X. Astron. Astrophys. \textbf{483} 15 (2008).
\bibitem{Alcaniz} Alcaniz, J.S.  Dev, A.  Jain, D. Astrophys. J. \textbf{627} 26 (2005).
\bibitem{Dev} Dev, A.  Safonova, M.  Jain, D. Lohiya, D. Phys. Lett. \textbf{B 548}  12 (2002).
\bibitem{36}Hawking, S.W. and Ellis, G.F.R. 
 Cambridge University press, Cambridge.
\bibitem{37}Wald, R. M. 
 The University of Chicago Presss, Chicago (1984).

\bibitem{38}Visser, M. Barcelo, C.
 COSMO-99  \textbf{112}, 98 (2000).
\bibitem{39}Barcelo, C. Visser, M. 
 Int. j. Mod. Phys. \textbf{D 11}, 1553 (2002).

\bibitem{Ber07}Bertolami, O. Pedroa, F. G. and Le Delliou M.: Phys. Lett. \textbf{B 654} 165 (2007).

\bibitem{Kumar}Kumar, S. Mon. Not. R. Astron. Soc. \textbf{422} 2532–2538 (2012).

\bibitem{keon}Chakkrit Kaeonikhom, Burin Gumjudpai, Emmanuel N. Saridakis, Phys. Let. \textbf{B 695}  45–54 (2011).

\bibitem{rani}Sarita Rani,  Altaibayeva, A.  Shahalam, M.  Singha, J.K. JCAP \textbf{03} 031 (2015).

\bibitem{komatsu}  Komatsu, E. et al.,  Astrophys. J. Suppl. 180, 330-376 (2009).

\bibitem{hinshaw}  Hinshw, G. et al., Astrophys. J. Suppl. Ser. \textbf{208}, 19 (2013).
\end{thebibliography}
\end{document}